\documentclass[a4paper,fleqn]{cas-dc}
\usepackage{siunitx}
\usepackage[authoryear]{natbib}

\def\tsc#1{\csdef{#1}{\textsc{\lowercase{#1}}\xspace}}
\tsc{WGM}
\tsc{QE}

\begin{document}
\let\WriteBookmarks\relax
\def\floatpagepagefraction{1}
\def\textpagefraction{.001}

\shorttitle{}    

\shortauthors{}  

\title [mode = title]{A Unified Model for Highly Accurate ECG-Free Dynamic Coronary Roadmapping Using Spatio-Temporal Transformers}  

\tnotemark[1] 

\tnotetext[1]{} 

%

\author[1, 3]{Saahil Islam}

\cormark[1]

\fnmark[1]

\ead{saahil.islam@fau.de}

\ead[url]{}

\credit{}

\affiliation[1]{organization={Pattern Recognition Lab, Friedrich Alexander University},
            city={Erlangen},
            country={Germany}}

\author[2]{Sebastian Piat}
\author[2]{Venkatesh N. Murthy}
\author[2]{Serkan Cimen}
\author[2]{Puneet Sharma}
\author[1]{Andreas Maier}
\author[3]{Florin C. Ghesu}




\credit{}

\affiliation[2]{organization={Digital Technology and Innovation, Siemens Healthineers},
            city={Princeton},
            country={USA}}

\affiliation[3]{organization={Digital Technology and Innovation, Siemens Healthineers},
            city={Erlangen},
            country={Germany}}

\cortext[1]{Saahil Islam}

\fntext[1]{}


\begin{abstract}
Percutaneous Coronary Intervention (PCI) is a minimally invasive procedure designed to restore coronary blood flow obstructed by atherosclerotic plaque. During PCI, repeated injections of iodine-based contrast agents are required to visualize the coronary arteries and guide interventional devices. However, frequent injections increase radiation exposure and the risk of contrast-induced nephropathy, with acute kidney injury reported in up to 30\% of patients with renal impairment.
Dynamic Coronary Roadmapping (DRM) has emerged as an effective strategy to mitigate these risks by overlaying a precomputed angiographic vessel map onto live fluoroscopy and continuously updating it during the procedure. Accurate DRM depends on precise cardiac phase matching between angiography and fluoroscopy, as well as reliable catheter tip tracking for motion compensation. These tasks remain challenging in electrocardiogram-free settings or when manual annotations are limited.
In this work, we introduce a unified DRM framework that simultaneously performs cardiac phase matching and catheter tip tracking, enabling accurate and real-time guidance. Our approach employs a large-scale spatio-temporal encoder pretrained on 16 million frames to model cardiac motion dynamics. To our knowledge, it is the first time this kind of pretraining has been used for motion compensation in DRM. Furthermore, we propose auxiliary tasks based on ECG R-peak detection and catheter tip tracking, which stabilize optimization and eliminate the need for extensive catheter mask annotations. Finally, a majority-voting postprocessing strategy aggregates temporal predictions to enhance robustness and provides a confidence score that correlates with the phase-matching error.
Comprehensive evaluations on clinical X-ray datasets show that the proposed method achieves state-of-the-art performance, producing low temporal misalignment and consistent phase-matching accuracy suitable for real-time DRM applications.

\end{abstract}





\maketitle

\section{Introduction}\label{}
Percutaneous Coronary Intervention (PCI) is a minimally invasive cardiology procedure performed to restore blood flow in coronary arteries obstructed by atherosclerotic plaque deposits, a condition commonly referred to as stenosis (\cite{khan2022percutaneous}). Under X-ray angiographic guidance, a catheter is navigated into the ostium of the coronary artery. Through this guiding catheter, a balloon catheter equipped with a stent is advanced over a guidewire to the stenotic site. Inflation of the balloon expands the stent, which is subsequently deployed to maintain vessel patency and prevent restenosis. To visualize the coronary vasculature and guide the catheter and guidewire to the stenotic region, an iodine-based contrast agent is injected during the procedure. However, opacification of coronary arteries only lasts for a short period of time and the contrast agent is injected multiple times for navigation of devices. Moreover, cardiologists have to mentally reconstruct the position of vessels and stenosis based on previous angiograms. This practice poses risks such as radiation exposure and contrast-induced nephropathy. In patients with pre-existing renal impairment, the incidence of acute kidney injury can be as high as 30\% (\cite{piayda2018dynamic, tehrani2013contrast}). 

\subsection{Dynamic Coronary Roadmapping}

Dynamic Coronary Roadmapping (DRM) has emerged as a promising technique to mitigate these risks by reducing both the radiation dose and the use of contrast agents (\cite{elion1989dynamic, zhu2010image, manhart2011self, kim2018registration, piayda2018dynamic, ma2020dynamic, liu2024auxiliary}). DRM leverages a detailed coronary map precomputed from angiography, which is superimposed on live fluoroscopy. The map is continuously updated in real time with each acquired fluoroscopic frame during PCI, providing cardiologists with immediate visual feedback throughout the intervention. This facilitates accurate guidewire navigation to the appropriate coronary branch and ensures precise placement of the stent at the stenotic site, while simultaneously reducing patient exposure to radiation and contrast agents.

Developing a DRM system necessitates the precise superimposition of a precomputed coronary artery map onto live fluoroscopic images. This process is particularly challenging because fluoroscopy provides only limited vessel information, complicating the compensation of motion caused by both cardiac activity and respiration. During the cardiac cycle, the coronary vessel tree undergoes dynamic deformation due to myocardial contraction and relaxation, requiring the selection of an anatomically appropriate vessel configuration for accurate mapping. In comparison, respiratory motion can often be approximated as translational; however, it still introduces additional displacement on top of cardiac-induced motion, thereby further complicating the alignment process.

\begin{figure}[]
  \centering
    \includegraphics[width=0.5\textwidth]{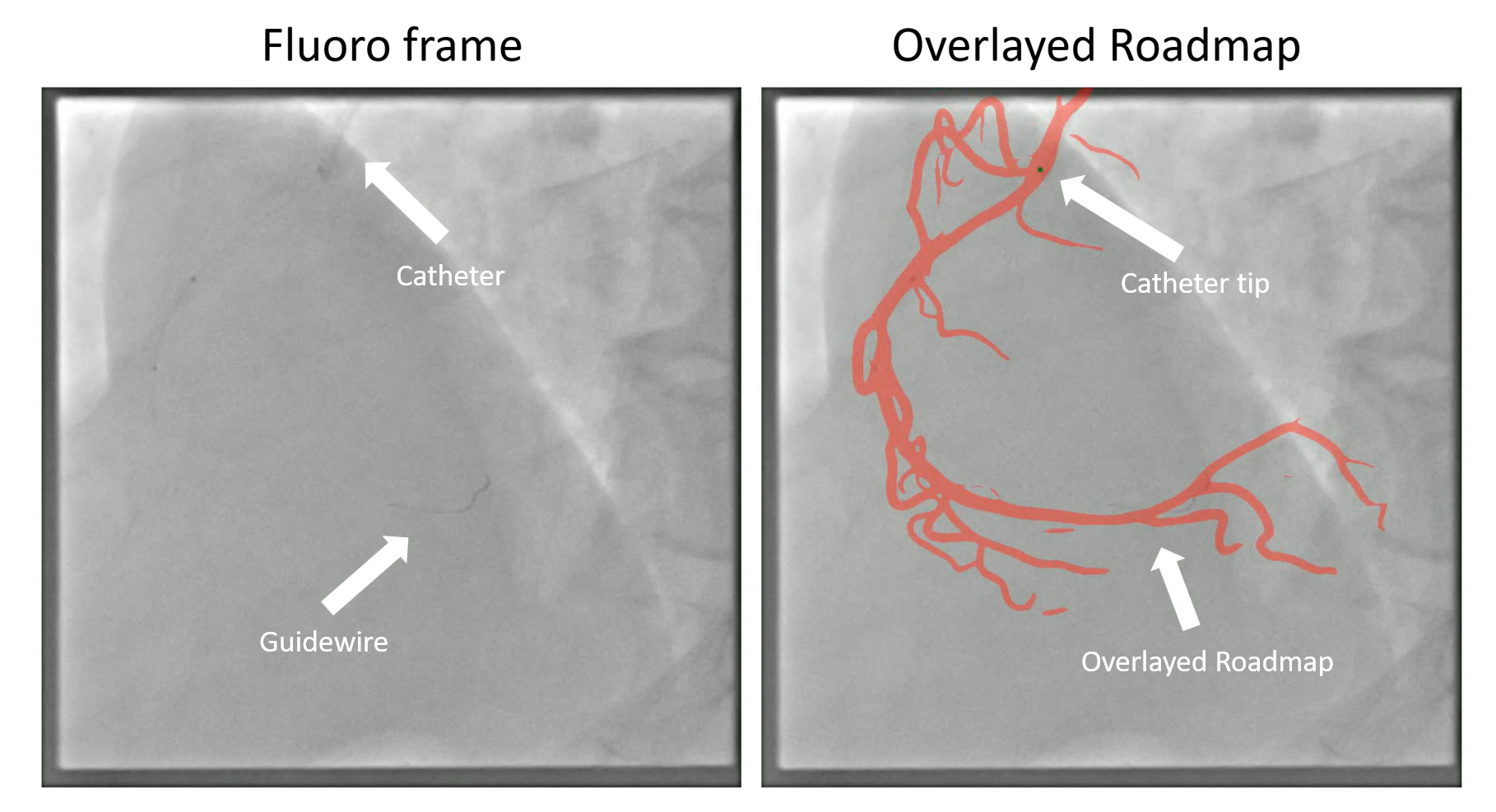}
    \caption{Example of dynamic coronary roadmap. The image on the left shows a fluoroscopy frame and the image on the right shows the overlayed vessel roadmap on the fluoroscopy frame. An accurate overlayed roadmap typically has the guidewire under the correct vessel branch.}\label{fig:drmeg}
\end{figure}

The earliest DRM system, proposed by \cite{elion1989dynamic}, generated roadmaps through digital subtraction of contrast-enhanced and mask sequences spanning a full cardiac cycle. These roadmaps were later synchronized with live fluoroscopy by aligning R-waves in the ECG. Although this method accounted for cardiac motion, it did not address respiratory motion. Later works, such as \cite{zhu2010image} and \cite{manhart2011self}, introduced image-based respiratory compensation methods. These approaches assumed an affine respiratory motion model in ECG-gated fluoroscopic frames and estimated displacement from soft-tissue motion while handling static structures separately. Their effectiveness, however, depends on sufficient visibility of relevant tissue in the field of view and is limited to cardiac-gated frames.  

\cite{kim2018registration} proposed creating binary vessel masks from at least one cardiac cycle of angiographic images to serve as roadmaps. Cardiac motion was compensated by temporally aligning angiographic roadmaps with fluoroscopy via cross-correlation of ECG signals, while respiratory motion was corrected by aligning the guidewire centerline in fluoroscopy with vessel contours from angiography. Although this method demonstrated feasibility in phantom studies, it lacked quantitative evaluation of spatiotemporal accuracy and relied heavily on robust vessel and guidewire extraction, which remains challenging in X-ray images. 

In contrast to direct roadmapping, model-based approaches predict motion in fluoroscopic frames using surrogate signals such as ECG-derived cardiac motion or respiratory motion from diaphragm tracking. Some studies model both cardiac and respiratory motion (\cite{shechter2005prospective, timinger2005motion, faranesh2013integration, fischer2017mr}), while others focus solely on respiratory motion using cardiac-gated images (\cite{schneider2010model, king2009subject, peressutti2013novel}). A major limitation of these methods is their patient-specific nature, requiring model retraining for each subject. Moreover, when surrogate values at inference fall outside the training range (e.g., due to abnormal motion), extrapolation is needed, which may compromise motion compensation accuracy.

Recent approaches employ deep learning for respiratory motion compensation, often by tracking the tip of the guiding catheter within a Bayesian filtering framework (\cite{ma2020dynamic}). However, this method is restricted to fluoroscopy and relies on manual detection in angiographic sequences, where catheter tracking is particularly challenging due to occlusions from contrast injection. To address this, \cite{demoustier2023contrack} proposed an optical flow-based solution, achieving superior tracking results compared to existing tracking methods in natural imaging (\cite{yan2021learning, cui2022mixformer, li2018high}). 
\cite{islam2024self, islam2024novel} leveraged pretrained transformers for improved robustness. Although these techniques demonstrate promising results for catheter tracking in both angiography and fluoroscopy, they have not been evaluated in the context of DRM. Furthermore, they require a manual initialization to achieve high robustness limiting full automation. 

Despite advances in respiratory motion correction, cardiac motion compensation in these methods still depends on ECG signals, necessitating ECG acquisition during both angiography and live fluoroscopy. \cite{liu2024auxiliary} show that cardiac phase matching can be achieved using image-based features obtained from deep-learning based feature extractors for angiography and fluoroscopy alignment. However, the network struggles to learn and converge unless a large dataset is used containing catheter mask annotations. This is due to the methodological limitation of encoder typically focusing on first image feature extraction and temporal fusion on highly downsampled image features. Hence, the motion cue is introduced into the model from the motion of the catheter masks. Obtaining such annotations for both angiography and fluoroscopy is highly labor-intensive and costly. Moreover, their framework relies on separate models for catheter tip tracking and phase matching, which limits its efficiency for real-time DRM. 

\subsection{Spatio-temporal features and feature matching}
Different strategies have been developed to learn spatio-temporal representations from both labeled and unlabeled data, demonstrating strong effectiveness in tasks where temporal dynamics play a important role. Most existing methods rely on self-supervised learning from large-scale unlabeled datasets and have been predominantly applied to action recognition in natural images (\cite{tong2022videomae, wang2023videomae, bardes2024revisiting, assran2025v}). However, while action recognition models capture long-range temporal dependencies, they often overlook subtle inter-frame variations. In contrast, approaches such as SAM2 (\cite{ravi2024sam}) achieve fine-grained frame-level tracking and segmentation but require extensive labeled data for training. On the other hand, approaches such as dino (\cite{caron2021emerging, oquab2023dinov2}) are pretrained only on images to learn spatial features and performs well on video segmentation tasks, it lacks consistency over the frames and fails during object occlusions. FIMAE (\cite{islam2024self}), trained in a self-supervised manner on unlabeled angiography and fluoroscopy datasets, addresses these limitations by learning representations sensitive to fine inter-frame correspondences. Furthermore, HiFT (\cite{islam2024novel}) enhances the learned feature space by incorporating auxiliary supervision from vessel segmentation tasks on annotated data.

In parallel, contrastive and metric learning approaches aim to structure the embedding space by bringing similar (positive) examples closer while pushing dissimilar (negative) examples apart (\cite{schroff2015facenet, khosla2020supervised, chen2020simple, he2020momentum, caron2020unsupervised}). The classical contrastive loss (\cite{Hadsell1640964}) enforces this separation by minimizing the distance between positive pairs and ensuring negatives remain beyond a defined margin. While initially introduced for supervised learning (\cite{khosla2020supervised, ghojogh2020fisher}), contrastive learning has since become central to self-supervised representation learning. Triplet loss extends this concept by operating on triplets $(a, p, n)$—an anchor $a$, a positive $p$ (similar or same class), and a negative $n$ (dissimilar or different class)—and constraining the embedding space such that the anchor-positive distance is at least a margin smaller than the anchor-negative distance (\cite{schroff2015facenet, zhao2019weakly}). This formulation was recently employed by (\cite{liu2024auxiliary}) for cardiac phase matching in DRM, where frames of the same cardiac phase are pulled closer in the embedding space, while those of different phases are pushed apart. 

\subsection{Contributions}
Despite recent progress, existing approaches for DRM face several major limitations. Many methods rely on ECG signals for cardiac phase alignment, which are not always available or reliable in clinical practice. Image-based approaches, on the other hand, often require large amounts of dense annotations, such as catheter masks, to provide sufficient motion cues for learning. Furthermore, prior works typically address cardiac phase matching and catheter tip tracking using separate models, leading to increased computational complexity and limiting real-time applicability. In addition, differences in image quality between angiography and fluoroscopy, as well as motion inconsistencies caused by respiration, patient movement, or device manipulation, further complicate robust alignment. These challenges highlight the need for a unified and annotation-efficient framework that can generalize across diverse imaging conditions.

To address these challenges, we propose a unified framework for DRM that simultaneously addresses cardiac phase matching and catheter tip tracking for accurate and real-time guidance. Our main contributions are summarized as follows:
\begin{itemize}
    \item We use a pretrained spatio-temporal encoder trained on 16 million frames for cardiac phase matching. To the best of our knowledge, this is the first work to leverage spatio-temporal pretrained features for cardiac motion compensation, achieving superior phase-matching performance compared to existing methods.
    \item We propose a novel pipeline that employs ECG R-peak detection and catheter tip tracking as auxiliary tasks, enabling stable training and convergence of phase matching without requiring additional catheter mask annotations.
    \item We design a unified model that jointly performs phase matching and catheter tip tracking, resulting in a fast inference suitable for real-time DRM.
    \item We develop a majority-voting post-processing strategy that incorporates predictions from previous fluoroscopy frames, enhancing robustness and producing a reliable confidence score. Through quantitative analysis, we demonstrate how this score correlates with the phase-matching error, making it valuable in interventional settings.
    \item We provide both visual and quantitative evaluations on clinical angiography and fluoroscopy sequences, demonstrating low errors for DRM and accurate catheter tip tracking.
\end{itemize}

\section{Methods}
In this section, we explain our proposed methodology for training and inference.
\subsection{Training}

\begin{figure*}
  \centering
    \includegraphics[width=\textwidth]{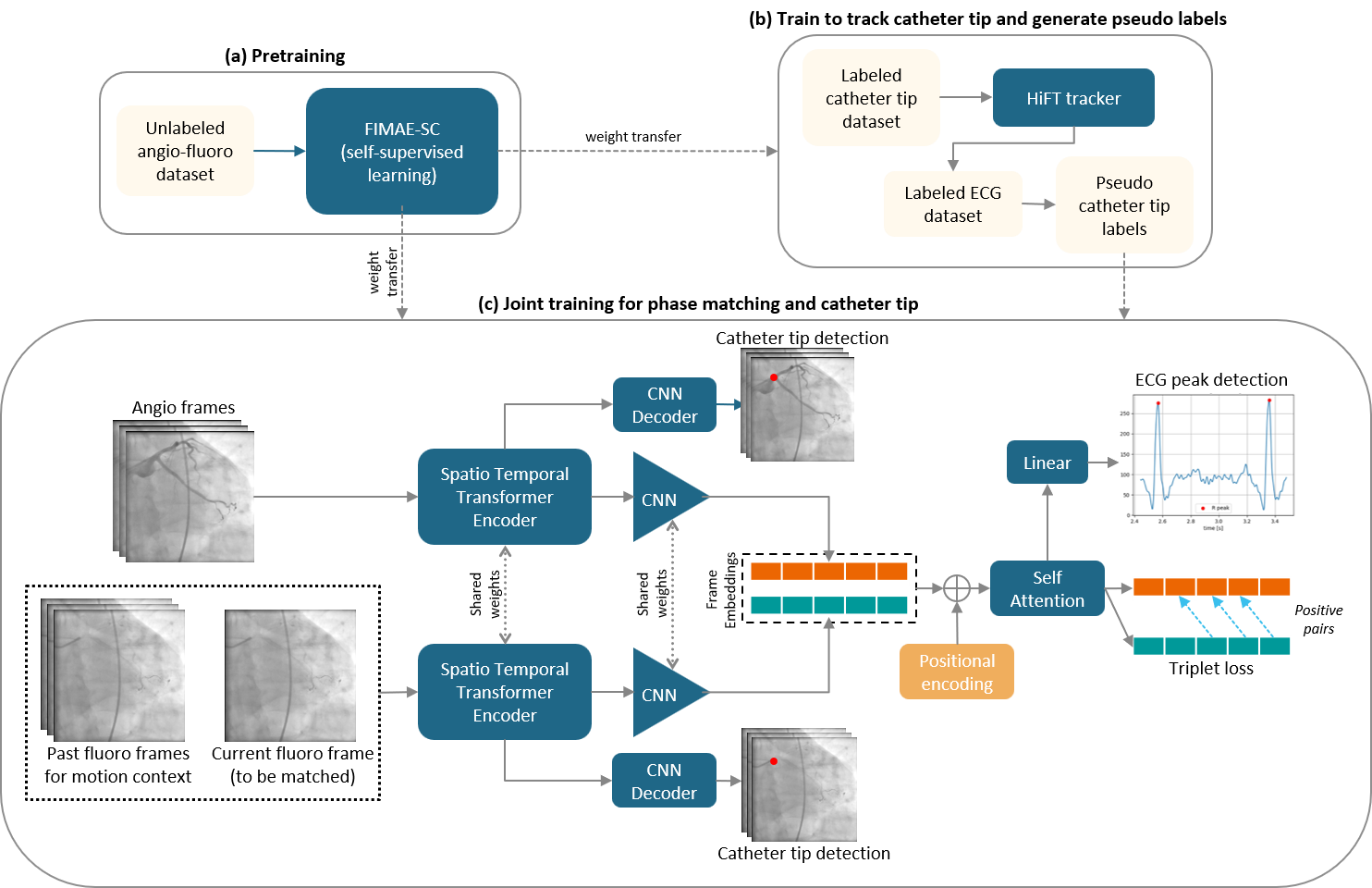}
    \caption{Overview of our method: We train our model in 3 stages; (a) First a spatio-temporal transformer encoder is pretrained on a large unlabeled dataset of angiography and fluoroscopy, (b) The pretrained encoder is used to train a model to track the catheter tip and the trained model is then used to generate pseudo catheter tip labels for a labeled ECG dataset. (c) Finally, a unified DRM model is trained with the pretrained backbone jointly on the ECG labels and pseudo catheter tip labels.}\label{img_drmtraining}
\end{figure*}

The overview of our training pipeline is shown in Fig.~\ref{img_drmtraining}.  
Given a live fluoroscopy frame $I^F$, our objective is to identify the corresponding frame in the angiography sequence, $S_A = [I^A_1, I^A_2, \dots, I^A_n]$ that matches its cardiac phase. Since a single fluoroscopy frame lacks sufficient information to capture cardiac motion, we instead consider a fluoroscopy sequence  $S_F = [I^F_1, I^F_2, \dots, I^F_n]$, assuming the last frame is the one we are concerned with matching and the other (past) frames are responsible for the motion context. 

We employ a pretrained spatio-temporal encoder $\mathbf{E}_\theta$, trained in a self-supervised manner on large-scale angiography and fluoroscopy datasets (\cite{islam2024self}), to extract features from both $S_A$ and $S_F$. Shared weights are used for the two modalities:  
\[
f^A = \mathbf{E}_\theta(S_A), \quad f^F = \mathbf{E}_\theta(S_F).
\]  

Note that since the pretrained network was trained with $10$ frames, we limit $n \leq 10$.
To reduce memory consumption during training, we alternately stop gradient flow for either angiography or fluoroscopy feature extraction. The spatio-temporal features $f^A, f^F$ are then processed by a lightweight CNN $\mathbf{C}_\phi$ (4 convolutional layers with max pooling followed by global average pooling), producing 768-dimensional frame-level embeddings:  
\[
\hat{e}^A = \mathbf{C}_\phi(f^A), \quad \hat{e}^F = \mathbf{C}_\phi(f^F).
\]  

Finally, embeddings from all frames are concatenated along the temporal dimension and combined with learnable positional encodings $\mathbf{P}$ to distinguish between angiography and fluoroscopy frames and to provide temporal context for each frame. These positional encodings are distinct from those used in the pretrained encoder and are learned specifically for the joint sequence. In particular, separate sets of positional embeddings are maintained for angiography and fluoroscopy, with a maximum capacity of 10 embeddings per modality. For a given input, only the required number of embeddings is selected (e.g., 5 out of 10 for a 5-frame sequence), allowing the model to encode both the temporal order and the modality-specific identity of each frame. We apply multi-headed self-attention layers on joint frame embeddings, $\hat{e}$:  
\[
\hat{e} = \text{Concat}([\hat{e}^A, \hat{e}^F]) + \mathbf{P}.
\]

This step is followed by multi-headed self-attention (MHA) layers on the joint embeddings for the model to learn the motion relationship between the angiography and fluoroscopy frame embeddings as well as for the model to learn a unique embedding for each cardiac phase in order to match between the sequences. For each attention head $h$, queries, keys, and values are computed as  
\[
Q_h = \hat{e}W_h^Q, \quad K_h = \hat{e}W_h^K, \quad V_h = \hat{e}W_h^V,
\]  
where $W_h^Q, W_h^K, W_h^V \in \mathbb{R}^{d \times d_h}$ are learnable projection matrices, $d$ is the embedding dimension of the input features, and $d_h$ is the embedding dimension of a single attention head.  
The attention output for head $h$ is defined as  
\[
\text{Attention}(Q_h, K_h, V_h) = \text{Softmax}\!\left(\frac{Q_h K_h^\top}{\sqrt{d_h}}\right)V_h.
\]  

The outputs of all $H$ heads are concatenated and linearly projected to obtain the final representation, $e = \text{MHA}(\hat{e})$. Although cross-attention is often used in matching tasks, our early experiments showed that self-attention performed better than cross-attention.  

As noted in \cite{liu2024auxiliary}, the loss landscape for cardiac phase matching is unsmooth, since frame-to-frame motion is minimal and a full cardiac cycle may not be visible when the encoder is limited to $10$ frames. To aid convergence and avoid local minima, we introduce an auxiliary task in which the network predicts the R-peaks of the ECG signal. For this, $e$ is passed through a linear projection followed by a sigmoid activation, yielding a scalar output $\hat{y} \in [0,1]$ indicating whether a frame corresponds to an R-peak ($1$) or not ($0$).  

Since R-peaks are sparse (often none or at most a few in a sequence), we use an adaptive weighted binary cross-entropy loss that balances positive and negative classes dynamically:  
\[
\mathcal{L}_{R_p} = - \frac{1}{N} \sum_{i=1}^N w_i \Big[ y_i \log \sigma(z_i) + (1-y_i)\log(1-\sigma(z_i)) \Big],
\]  
where $z_i$ are the logits, $y_i \in \{0,1\}$ are the ground-truth labels, $\sigma(\cdot)$ is the sigmoid function, and $w_i$ is the adaptive weight.

Let $N_0$ and $N_1$ denote the number of negative and positive samples in the batch, respectively, with $N=N_0+N_1$. The adaptive weights are defined as  
\[
w_i = 
\begin{cases}
\frac{N}{2N_0}, & y_i = 0, \\[6pt]
\frac{N}{2N_1}, & y_i = 1,
\end{cases}
\]  
ensuring equal contribution from positive and negative samples regardless of class imbalance.

For the cardiac phase matching, we employ a triplet loss on the attended embeddings $e$.  
Given an anchor embedding $e^F_a$ from a fluoroscopy frame, a positive embedding $e^A_p$ from an angiography frame of the same cardiac phase, and a negative embedding $e^A_n$ from a different phase, the triplet loss is defined as:  
\[
\mathcal{L}_{P_m} = \max\Big(0, \; d(e^F_a, e^A_p) - d(e^F_a, e^A_n) + \alpha \Big),
\]  
where $d(\cdot,\cdot)$ is cosine similarity and $\alpha$ is the margin with a value of $0.8$.

Finally, we introduce catheter tip tracking as an auxiliary task, making the network unified for all the requirements of dynamic coronary roadmapping. This task provides additional cues for phase matching by enforcing the network to learn catheter motion. Since only a few frames per sequence are annotated (3–5 on average), we employ Historical Feature Guided Tracker (HiFT) \cite{islam2024novel} to generate pseudo-labels for all frames.  

For both angiography and fluoroscopy sequences, the spatio-temporal features $f^A$ and $f^F$ are passed through a CNN-based upsampling head $\mathbf{C}_\Phi$. The outputs are concatenated and projected to obtain the predicted heatmap $\hat{H}$. We supervise $\hat{H}$ with pseudo-label heatmaps $H$ using Dice loss:  
\[
\mathcal{L}_{C_t} = 1 - \frac{2 \sum_{i} \hat{H}_i H_i}{\sum_{i} \hat{H}_i^2 + \sum_{i} H_i^2}.
\]  

The final loss for training is given by:
\[
\mathcal{L} = \lambda_{R_p}\mathcal{L}_{R_p} + \lambda_{P_m}\mathcal{L}_{P_m} + \lambda_{C_t}\mathcal{L}_{C_t}
\] 

Where $\lambda_{R_p}$, $\lambda_{P_m}$ and $\lambda_{C_t}$ are loss weights chosen as $0.5$, $1.0$ and $0.5$ respectively.

\subsection{Inference}
The inference procedure is divided into an \textit{offline phase} and an \textit{online phase}.  

\begin{figure}[]
  \centering
    \includegraphics[width=0.5\textwidth]{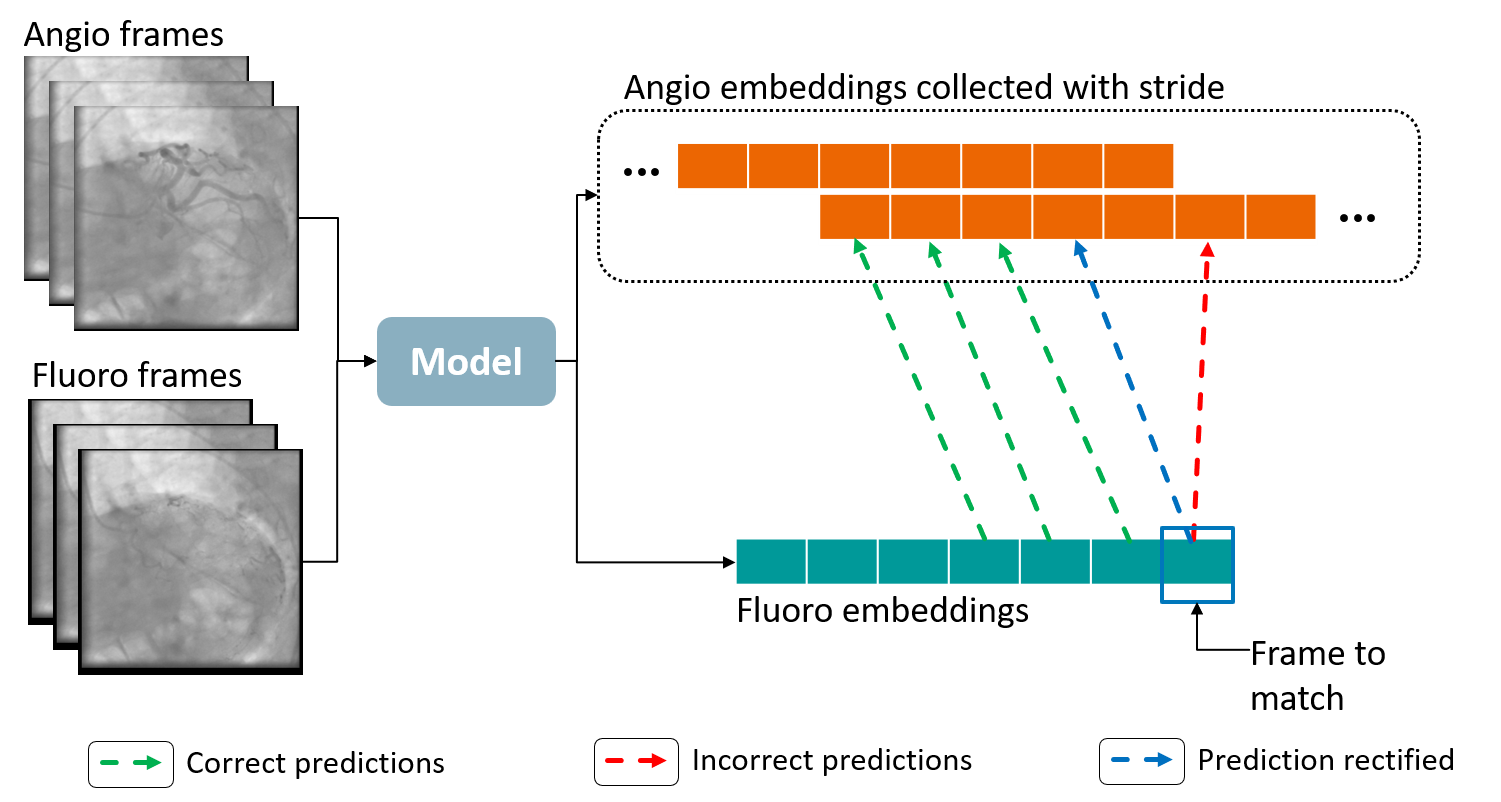}
    \caption{Proposed inference strategy with majority voting post-processing}\label{img_drminf}
\end{figure}

\subsubsection{Offline phase}
In the offline stage, we first apply a contrast detection model together with R-peak detection to identify one cardiac cycle (peak-to-peak) of highly contrasted frames. To construct the angiography sequence, an offset of two frames is applied. Since the network is limited to processing at most $10$ frames, we use a temporal window of $10$ frames with a stride of $4$ as the angiography sequence input. For each input sequence, all angiography frame embeddings $\hat{e}^A$ are precomputed and stored. A Res-UNet trained for vessel segmentation is used to extract vessel roadmaps, which are also stored for later use.  

\subsubsection{Online phase}
During the online phase, real-time inference is performed on incoming fluoroscopy frames. We employ a sliding temporal window of $10$ frames with a stride of $1$, where the last frame in the sequence is treated as the current live frame. For each fluoroscopy sequence, frame embeddings $\hat{e}^F$ are computed and concatenated with every stored angiography embedding $\hat{e}^A$. The embeddings are added with their respective positional encodings $\mathbf{P}$ and passed MHA layers, consistent with the training setup. Cosine similarity is then computed between fluoroscopy and angiography embeddings across all stored sets.  

To determine the final matching index, we employ a majority-voting post-processing strategy. Specifically, for each fluoroscopy frame embedding, the angiography frame index with maximum cosine similarity is retrieved from every stored set. The matched indices are interpolated to estimate the most likely corresponding index for the last fluoroscopy frame in the sequence. The final index for the current live frame is then obtained via majority voting over all interpolated indices. An illustration of this majority-voting post-processing is shown in Fig.~\ref{img_drminf}.

\section{Experiments}
\subsection{Dataset and training details}
\begin{table*}[]
\caption{Comparison of different strategies and models for automatic image-based phase matching for dynamic coronary roadmapping. Angio-Fluoro refers to our internal large unlabeled dataset of angiography and fluoroscopy dataset and LVD-142M (\cite{oquab2023dinov2}) refers to their curated dataset consisting of 142 million natural images, used to train Dinov2.}
\label{tab:phasematch}
\begin{tabular}{cccccccc}
\hline
Strategy              & Pretrained   & Feature Extraction           & \multicolumn{3}{c}{\makecell{Distance error \\ (Frames)}} & \makecell{Distance error \\ (Time in ms)} & \makecell{Distance error \\ (Percentage)} \\ \hline
                      &              &                 & mean $\pm$ std        & median      & max       & mean $\pm$ std                  & mean $\pm$ std                  \\ \hline
\makecell{AIT \\ (\cite{liu2024auxiliary})}                   & ImageNet     & CNN-Transformer & 1.20 $\pm$ 0.73        & 1.0         & 5.0       & 98.51 $\pm$ 68.35                      & 7.67 $\pm$ 4.25                      \\ \hline
\multirow{5}{*}{Ours} & ImageNet     & CNN-Transformer & 1.02 $\pm$ 0.83       & 0.77        & 3.69      & 84.85 $\pm$ 76.78                      & 6.65 $\pm$ 5.49                        \\
& LVD-142M & Dinov2-Transformer & 1.08 $\pm$ 1.01 & 0.91 & 3.40 & 89.38 $\pm$ 85.14  & 8.48 $\pm$ 7.0 \\
                      & Angio-Fluoro & Dinov2-Transformer          & 0.94 $\pm$ 0.68       & 0.81        & 3.45      & 77.8 $\pm$ 63.81                       & 6.19 $\pm$ 4.83                       \\ 
                      & None         & Spatio-Temporal        & 2.78 $\pm$ 1.25       & 2.7         & 7.0       & 207.56 $\pm$ 95.67                      & 16.70 $\pm$ 7.8                      \\ 
                      & Angio-Fluoro & Spatio-Temporal       & \textbf{0.64 $\pm$ 0.66}       & \textbf{0.55}        & \textbf{3.15}      & \textbf{56.31 $\pm$ 62.43}                       & \textbf{4.38  $\pm$ 4.51}  \\ \hline                   
\end{tabular}
\end{table*}

\begin{figure}[]
  \centering
    \includegraphics[width=0.5\textwidth]{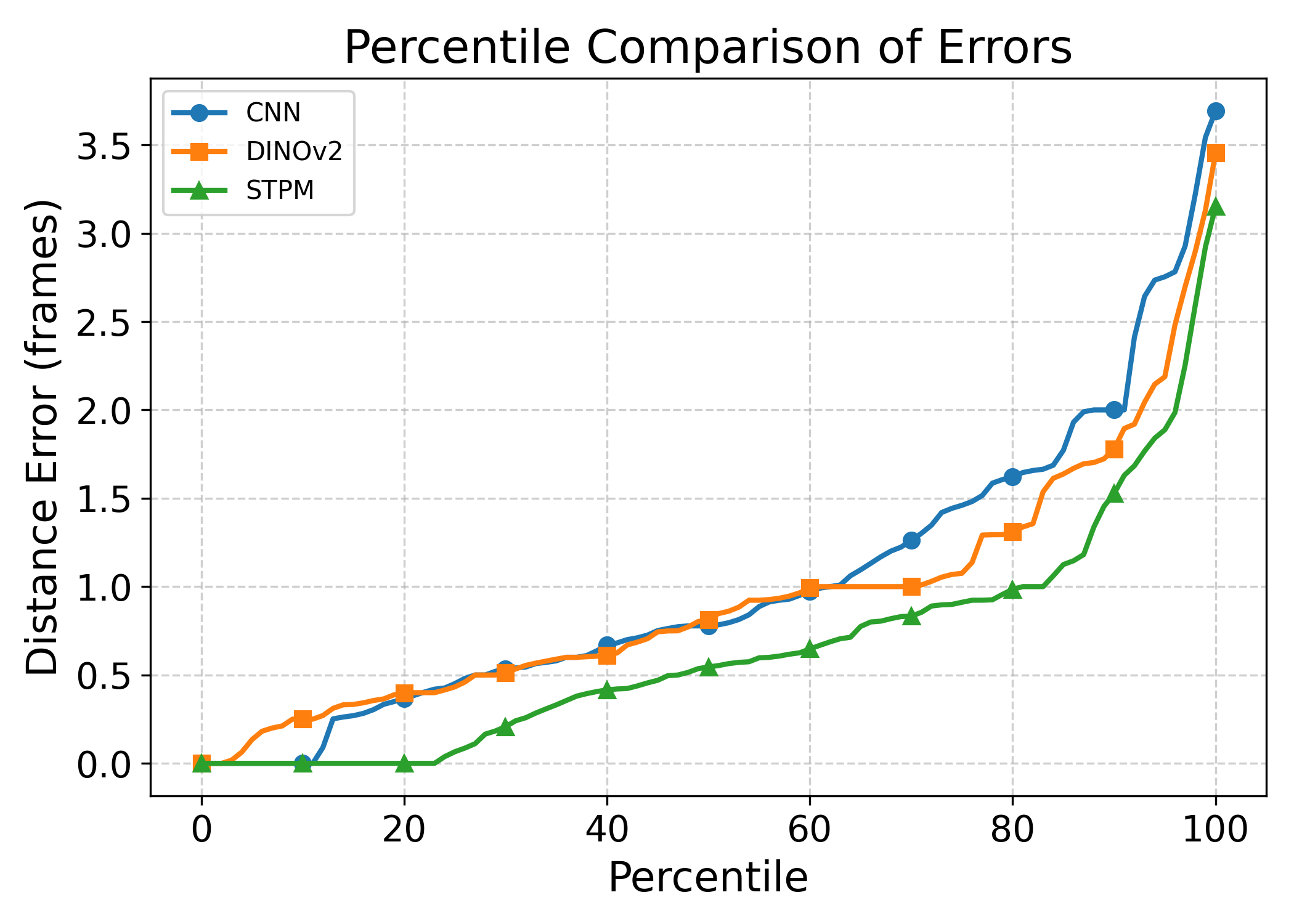}
    \caption{Percentile distribution of phase-matching errors and comparison of different backbones with our proposed strategy. STPM refers to spatio-temporal phase matching (ours). }\label{fig:percentile_pm}
\end{figure}

Our phase-matching dataset comprises 1,434 training pairs of angiography and fluoroscopy sequences, totaling 116,256 frames. The validation set includes 204 sequences with 16,223 frames, while the test set contains 79 pairs with 6,676 frames, of which 1,404 are fluoroscopy frames. Each sequence is associated with synchronized ECG data, from which R-peaks are identified. To represent cardiac phase continuity, we assign a normalized phase value of $0$ to R-peak frames, with intermediate frames linearly interpolated between $0$ and $1$. When the number of frames between consecutive R-peaks differs across angiography and fluoroscopy sequences, the nearest phase value between the two is used to establish correspondence.

Additionally, we use a catheter tracking dataset comprising 2,314 training sequences with a total of 198,993 frames, among which 44,957 frames include catheter tip annotations.  A subset of this dataset contains catheter mask annotations ($\approx$ 10\% of the total number of frames). Since most sequences have sparse, non-consecutive annotations, we employ the pretrained HiFT model \cite{islam2024novel} to generate pseudo labels for unannotated frames. The pseudo catheter tip coordinates are converted to gaussian heatmap with a standard deviation of $\approx 2.5$mm serving as ground truth for catheter tip detection. This enables multi-task training for both phase matching and catheter tip tracking across all frames. The dataset containing paired ecg data is a subset of this larger dataset.

For pretraining the spatio-temporal encoder, we use an internal unlabeled coronary X-ray dataset similar to FIMAE-SC \cite{islam2024novel}. This dataset comprises 241,362 sequences collected from 21,589 patients, totaling 16,342,992 frames across both angiography and fluoroscopy modalities. Supplementary cues for FIMAE-SC pretraining are derived from a ResUNet trained on 3,300 angiography sequences (with 91 for testing), where coronary arteries were annotated with centerline points and approximate vessel radii for five highly contrasted frames to generate target vesselness maps.

All frames are resized to $512 \times 512$ and augmented with random affine transformations, including translation in the range $(-0.15, 0.15)$, rotation between $(-10^\circ, 10^\circ)$, scaling between $(0.8, 1.2)$, and random horizontal and vertical flips. The model is trained for 500 epochs using a cosine annealing scheduler with a linear warmup. The learning rate is initialized at \num{4e-6}, linearly increased for the first 30 epochs to \num{8e-6}, and then gradually decayed following a cosine schedule to \num{1e-7}.

The final model is chosen based on its phase-matching performance on the validation set. This model is then fine-tuned using catheter-tip labels generated by HiFT, during which only the CNN decoder for tip detection is updated while the remaining components are kept frozen. In preliminary experiments, training with the weighted loss introduced a substantial number of false positives, some of which overlapped with the true positives in the catheter-tip heatmaps, making it harder to remove with any kind of post-processing. These spurious responses were largely eliminated after the dedicated decoder fine-tuning stage.

\subsection{Results for phase matching}
We conduct a comprehensive evaluation of our method against state-of-the-art approaches, analyzing the effectiveness of the proposed strategy and further examining whether the model’s confidence scores exhibit a correlation with phase-matching errors in DRM. Consistent with the evaluation protocol of \cite{liu2024auxiliary}, we report distance errors in terms of both frames and frame percentage. To enhance interpretability and provide a more tangible understanding of temporal misalignment, we additionally express the distance errors in milliseconds. The complete results are summarized in Table~\ref{tab:phasematch}.

Overall, our approach attains state-of-the-art performance on the test dataset. The results demonstrate that the proposed strategy consistently surpasses AIT when employing the same backbone architecture (CNN-Transformer). In particular, our method achieves an average distance error of 56.31 milliseconds—corresponding to 4.38\% of the cardiac cycle—with a low standard deviation of 0.66, indicating highly stable and reliable phase-matching performance across diverse sequences. It is worth noting that only 10\% of the training dataset includes catheter mask annotations, making a direct comparison with \cite{liu2024auxiliary} less straightforward. Despite this, the findings underscore that integrating auxiliary tasks within the model provides sufficient supervisory signals, substantially mitigating the reliance on extensive manual catheter mask annotations.

Moreover, the results highlight the significance of jointly modeling spatial and temporal dynamics for effective phase matching. A unified spatio-temporal feature extractor consistently outperforms designs that decouple spatial encoding from temporal fusion. Finally, domain-specific pretraining on angiography and fluoroscopy data yields substantially superior performance compared to pretraining on unrelated datasets. This trend is evident when comparing DINOv2 pretrained on LVD-142M~\cite{oquab2023dinov2} with the same model pretrained on our internal large-scale dataset of angiography and fluoroscopy. Notably, DINOv2 pretrained on LVD-142M performs worse than the CNN-Transformer baseline, which can be attributed to the domain mismatch and the limited size of the downstream dataset. These findings further reinforce the importance of domain-adapted representation learning for achieving robust performance in DRM.

Figure~\ref{fig:percentile_pm} illustrates a percentile-wise comparison of phase-matching errors, measured in frame distance, across different models. The proposed Spatio-Temporal Phase Matching (STPM) approach consistently exhibits superior performance across nearly the entire distribution of test samples, achieving lower distance errors than both the CNN and DINOv2 baselines. These results indicate that STPM generalizes more robustly to diverse cardiac motion patterns and varying imaging conditions.

A notable observation is that STPM maintains a distance error below one frame for approximately 82\% of all test cases, underscoring its reliability and fine-grained temporal alignment capability. In contrast, DINOv2 and CNN reach this error threshold considerably earlier, reflecting a sharper degradation in more challenging frames. Moreover, for about 22\% of the samples, the error is exactly zero, signifying perfect frame-level phase alignment between angiography and fluoroscopy sequences. Beyond the 80th percentile, all models exhibit a sharper increase in error, which can be attributed to difficult imaging conditions such as extreme view angles that obscure cardiac motion or cases where the vessel tree in the angiography frame is partially cropped. Nevertheless, even within these challenging regions, STPM maintains the lowest error margin, demonstrating its ability to capture subtle inter-frame dynamics that other models fail to represent effectively.

\subsubsection{Confidence score and distance error}
\begin{figure}[]
  \centering
    \includegraphics[width=0.5\textwidth]{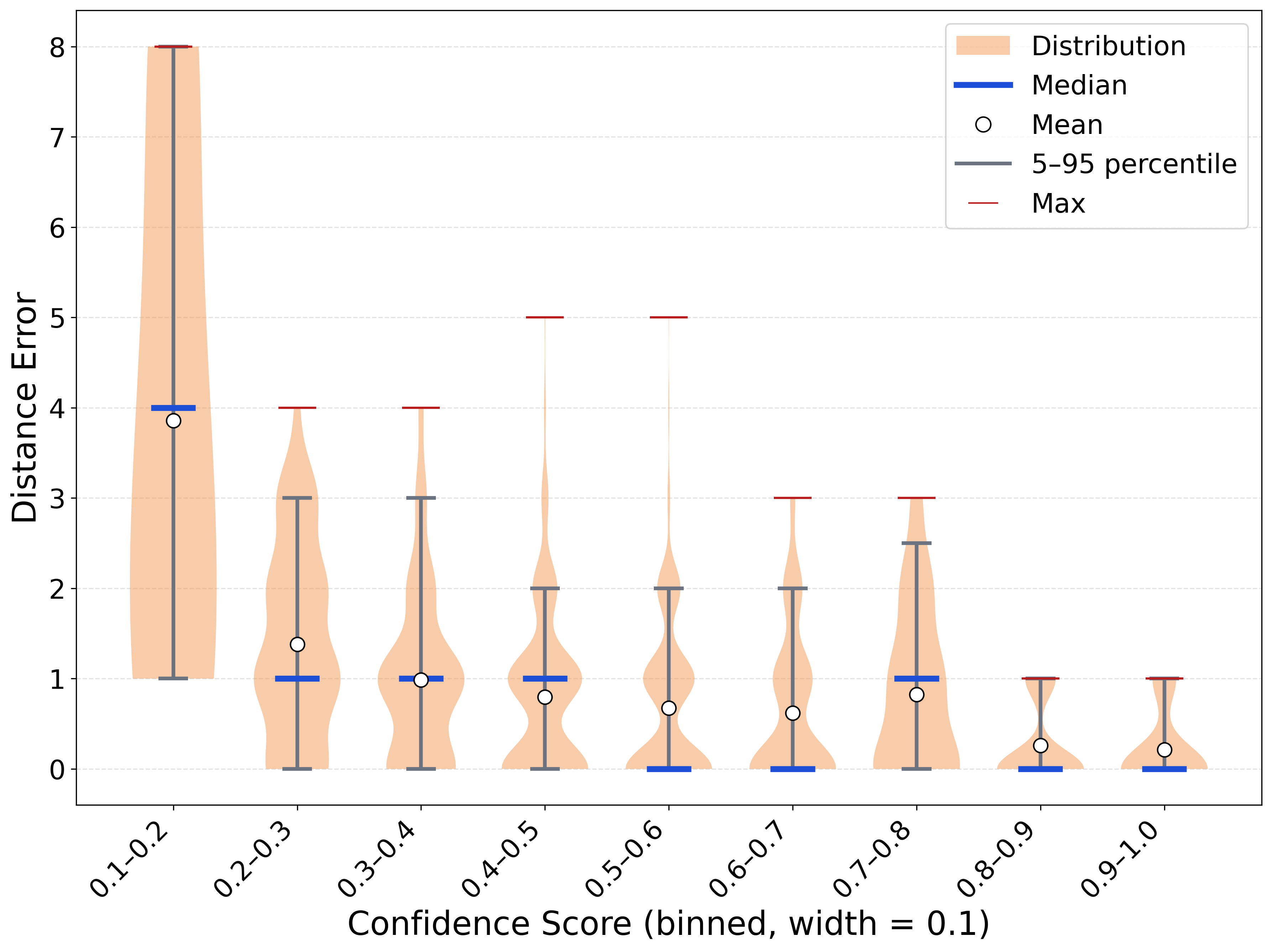}
    \caption{Distribution of error in each bins of confidence score obtained from the voting frequency of our model. Note that the distance error depicted here is in the frame level and minimum confidence score obtained in the test set was 0.16.}\label{fig:violin_confidence}
\end{figure}

\begin{figure}[]
    \centering
    \includegraphics[width=0.8\linewidth]{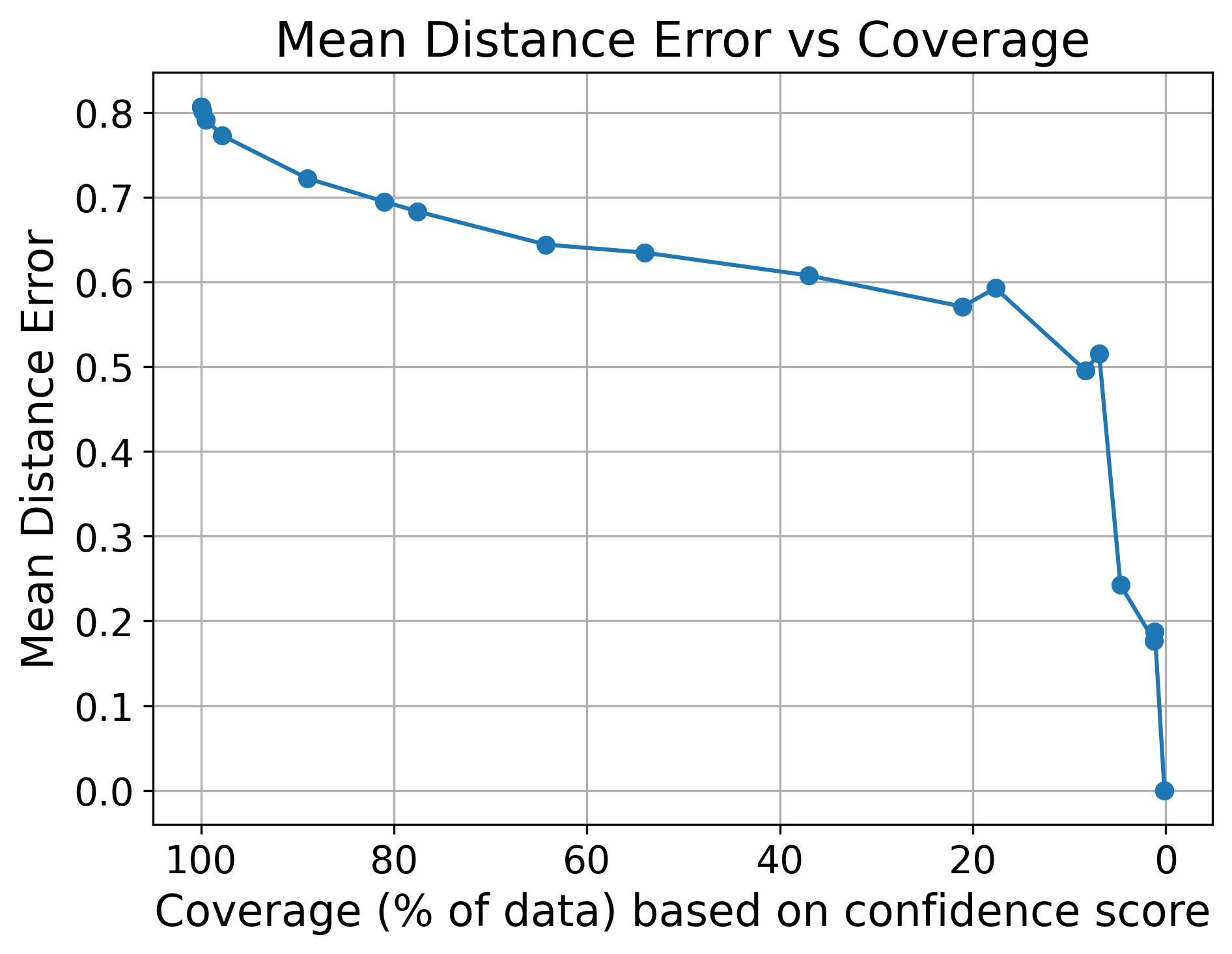}

    \vspace{0.6em}

    \includegraphics[width=0.8\linewidth]{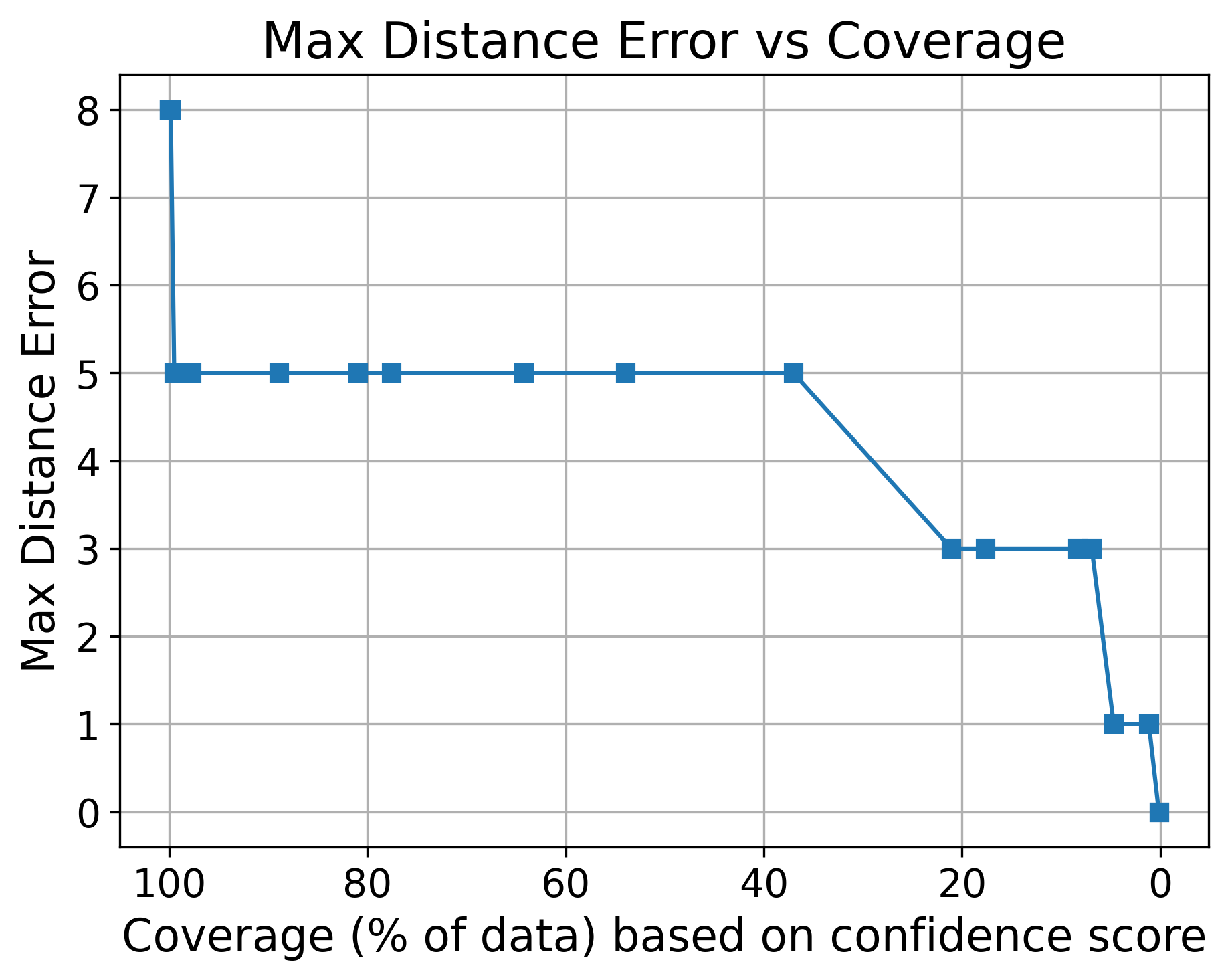}

    \caption{Distance error as a function of coverage (\% of data remaining for confidence score thresholds of 0.1).
    (Top) Mean distance error.
    (Bottom) Maximum distance error.  Note that the distance error depicted here is in the frame level.}
    \label{fig:error_vs_coverage}
\end{figure}

The majority-voting post-processing strategy employed during inference additionally enables the estimation of a confidence score for each prediction. This confidence score is defined as the ratio between the number of candidate predictions that agree on the same frame while being the majority and the total number of candidates considered for that prediction. In the context of image-guided therapy, the availability of an interpretable confidence measure is highly beneficial, as it allows surgeons or technicians to assess the reliability of the system’s output and appropriately weigh it against their own expertise. For such a confidence score to be meaningful, it is desirable that it exhibits an inverse relationship with the prediction error.

The relationship between the confidence score and the distance error is illustrated in Fig.~\ref{fig:violin_confidence} using a violin plot. Here, distance errors are evaluated at the frame level rather than being averaged over an entire sequence. The plot depicts the error distributions for confidence score bins of width 0.1. For confidence scores below 0.2, the error distribution is approximately uniform over the range of 1 to 8 frames, indicating that predictions with very low confidence are equally likely to be either close to the correct cardiac phase or to exhibit large errors. As the confidence score increases, a consistent reduction is observed in the mean, median, 5--95 percentile range, and maximum error. Notably, for confidence scores between 0.4 and 0.8, both the mean and median errors fall below one frame; however, isolated outliers with errors as large as five frames are still present. Beyond the overall reduction in error statistics, Fig.~\ref{fig:violin_confidence} also reveals how the variability of the predictions evolves with the confidence score. For low-confidence predictions, the error distributions are broad and exhibit substantial spread, indicating highly inconsistent behavior across samples. As the confidence score increases, these distributions become progressively narrower and increasingly concentrated around zero error. This suggests that the confidence score reflects not only the expected accuracy of a prediction but also its robustness, with higher confidence predictions exhibiting markedly reduced variability.

An interesting transition can be observed for confidence scores above approximately 0.8. In this regime, the error distributions collapse to near-zero values, with the median error reaching zero frames and the upper percentiles remaining tightly bounded. This indicates that the model is able to identify a subset of predictions for which the inferred cardiac phase is consistently accurate. Such behavior is particularly relevant in practical settings, where it may be preferable to rely only on predictions that the model itself deems reliable.

To further characterize predictive uncertainty, we analyze the relationship between distance error and data coverage as a function of the confidence score, as shown in Fig.~\ref{fig:error_vs_coverage}. Coverage is defined as the percentage of test samples retained after discarding predictions with confidence scores below a specified threshold. Starting from a threshold of 0.1, the threshold is incremented in steps of 0.1. A trend similar to the error distribution statistics over the confidence score is reflected in the coverage-based analysis. As predictions with lower confidence scores are progressively discarded, the mean distance error decreases in a near-linear manner, followed by a sharper drop when only the most confident predictions are retained. In contrast, the maximum error remains relatively high over a wide range of coverage values, indicating the presence of occasional large errors even when the average performance is reasonable. These outliers are largely removed only when the coverage falls below approximately 40\%, suggesting that confidence-based filtering is particularly effective at mitigating worst-case errors rather than merely improving average accuracy.

Overall, these observations indicate that the proposed confidence score is well aligned with the underlying prediction error. It provides a meaningful measure of uncertainty that can be used to balance accuracy and coverage, and offers a practical mechanism for identifying reliable predictions in live dynamic coronary roadmapping where erroneous outputs may have significant consequences.

\subsubsection{Ablation for phase matching}
\begin{figure}[]
  \centering
    \includegraphics[width=0.5\textwidth]{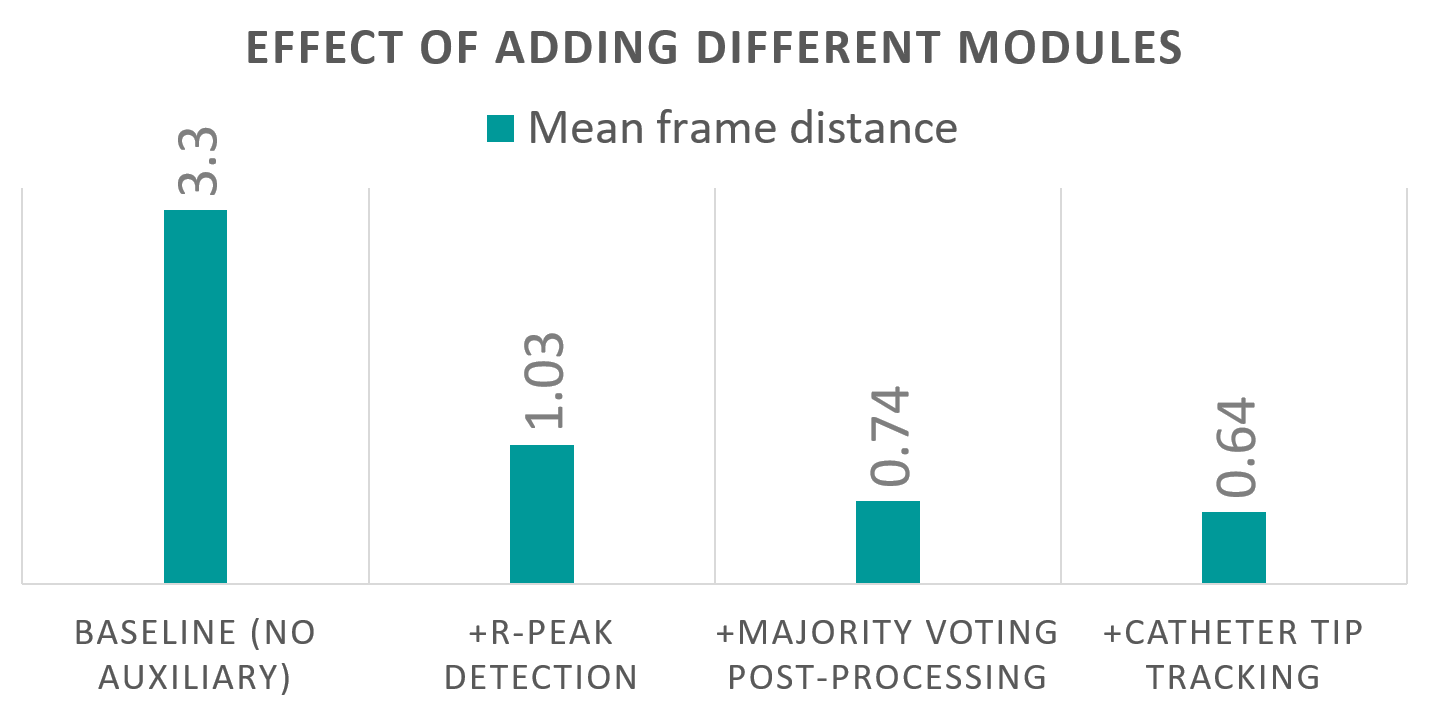}
    \caption{Effect of adding different modules in our spatio-temporal phase matching model}\label{fig:ablation}
\end{figure}

Figure~\ref{fig:ablation} presents the performance of our model under different architectural configurations, illustrating the contribution of each module within the pipeline. We observe that incorporating a simple R-peak detection auxiliary task enhances the model’s performance by 62.7\%. As noted in \cite{liu2024auxiliary}, the loss landscape associated with triplet loss in cardiac phase matching is inherently unsmooth. While \cite{liu2024auxiliary} address this issue by introducing auxiliary catheter mask inputs, our findings demonstrate that even a lightweight auxiliary task, such as ECG R-peak detection, is sufficient to stabilize training and facilitate convergence—thereby eliminating the need for extensive manual catheter mask annotations.

Moreover, integrating the proposed majority-voting postprocessing leads to a further and substantial improvement in performance. This gain is expected, as the aggregation of predictions across multiple candidates yields more robust and accurate outcomes. Finally, we observe that incorporating catheter tip tracking within the same model not only unifies the DRM framework but also yields additional gains in phase-matching accuracy, underscoring the mutual benefit of multi-task learning.

\subsection{Results for catheter tip tracking from unified model}
\begin{table}[]
\centering 
\caption{Comparison of the performance of different models for catheter tip tracking. Note that the difference in performance between HiFT and the unified model under automatic initialization is not statistically significant ($p > 0.1$).}
\label{tab:cathtip}
\begin{tabular}{c c c c c}
\hline
Init type & Model & \multicolumn{3}{c}{distance error (mm)} \\ \hline 
& & Mean $\pm$ std & Median & Max \\
\hline
\multirow{3}{*}{Manual} & ConTrack & 1.63 $\pm$ 1.70 & 1.08 & 13.32 \\
       & SimST    & 1.44 $\pm$ 1.35 & 1.02 & 10.23 \\
       & \textbf{HiFT} & \textbf{1.21 $\pm$ 0.68} & \textbf{1.04} & \textbf{4.04} \\
\hline
\multirow{5}{*}{Auto}   & AIT              & 1.91 $\pm$ 1.75 & 1.32 & 15.12 \\
       & ConTrack         & 2.87 $\pm$ 2.36 & 2.29 & 17.26 \\
       & SimST            & 2.24 $\pm$ 2.19 & 1.61 & 18.66 \\
       & \textbf{HiFT}    & \textbf{1.45 $\pm$ 1.30} & \textbf{1.05} & \textbf{9.29} \\
       & \makecell{Unified Model \\ (Ours)} & 1.58 $\pm$ 1.42 & 1.11 & 14.60 \\
\hline
\end{tabular}
\end{table}

Table~\ref{tab:cathtip} presents a quantitative comparison of catheter tip tracking performance across different models and initialization strategies. Most existing approaches rely on manual initialization of the catheter landmarks, which typically leads to improved accuracy but limits their applicability in a fully automated DRM workflow. As expected, methods evaluated under manual initialization consistently achieve lower mean and median errors compared to those operating with automatic initialization.

Among the manually initialized models, HiFT achieves the best overall performance, with substantially reduced maximum error compared to other approaches, indicating improved robustness in challenging cases. Under automatic initialization, performance degradation is observed across all methods, highlighting the difficulty of the task in the absence of manual guidance. In this setting, HiFT again attains the lowest errors, while the unified model achieves comparable performance with only a modest increase in mean error. It is worth noting that for ConTrack, SimST, and HiFT, a separate model trained with the same encoder as the tracker is used to detect the catheter tip in the first frame, serving as the initialization. In contrast, AIT and the proposed unified model perform both initialization and tracking within a single model, reducing the overall training time and eliminating the need for an additional initialization network.

The unified model exhibits slightly lower accuracy than the automatically initialized HiFT model, which is expected given that it is trained using pseudo-labels generated by HiFT due to the lack of frame-level manual annotations across the full training set. Notably, the difference in performance between the two models is not statistically significant ($p > 0.1$), suggesting that the observed gap is small. Moreover, the unified model maintains competitive median error while reducing reliance on model-specific initialization assumptions, making it better aligned with the requirements of a scalable and fully automated DRM pipeline.

\begin{figure}[]
  \centering
    \includegraphics[width=0.45\textwidth]{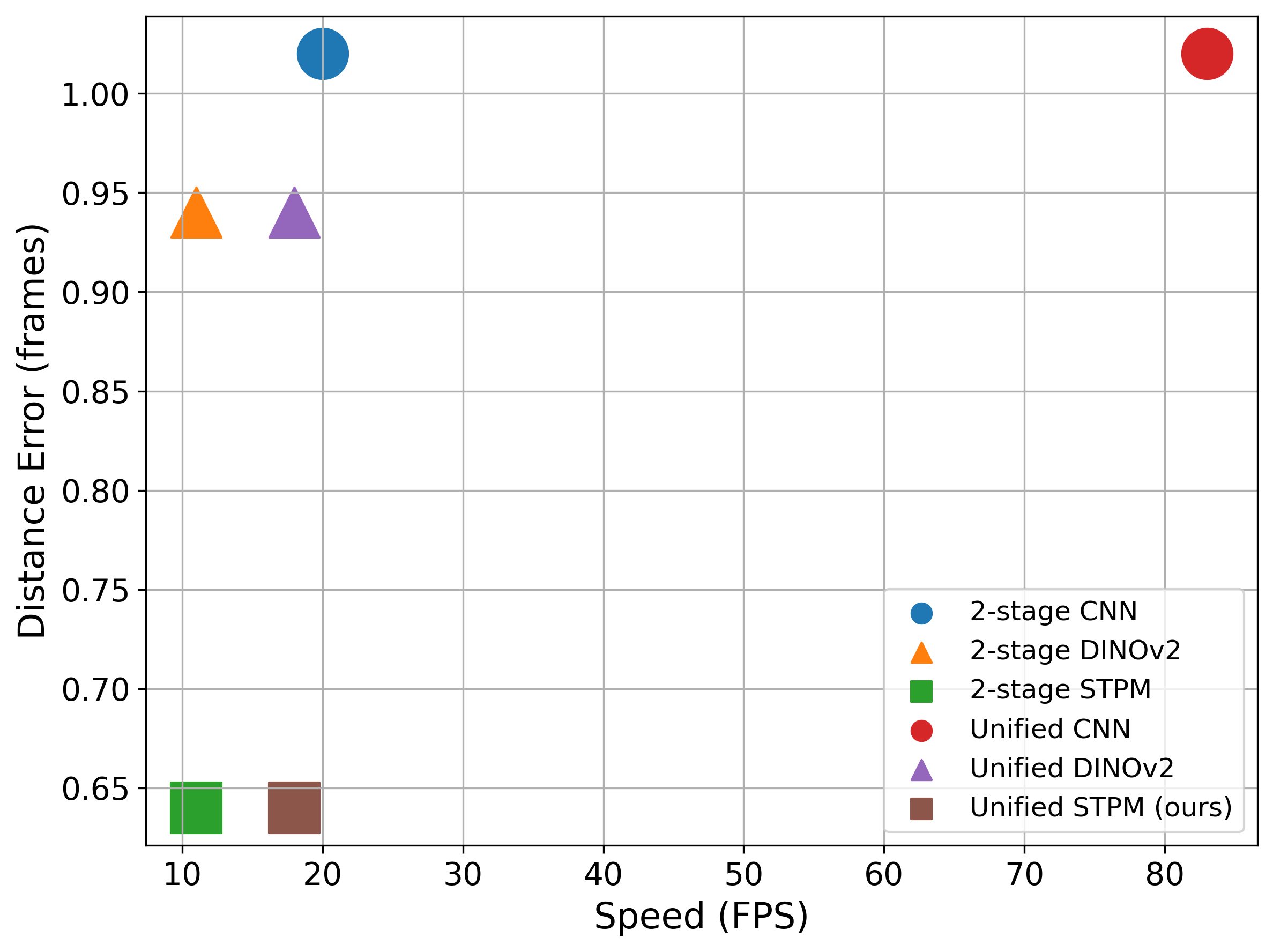}
    \caption{Comparison of model speed and phase matching error for different backbones in different setups}\label{fig:speed}
\end{figure}

\subsection{Phase matching error and inference speed}
\begin{figure*}[]
  \centering
    \includegraphics[width=0.75\textwidth]{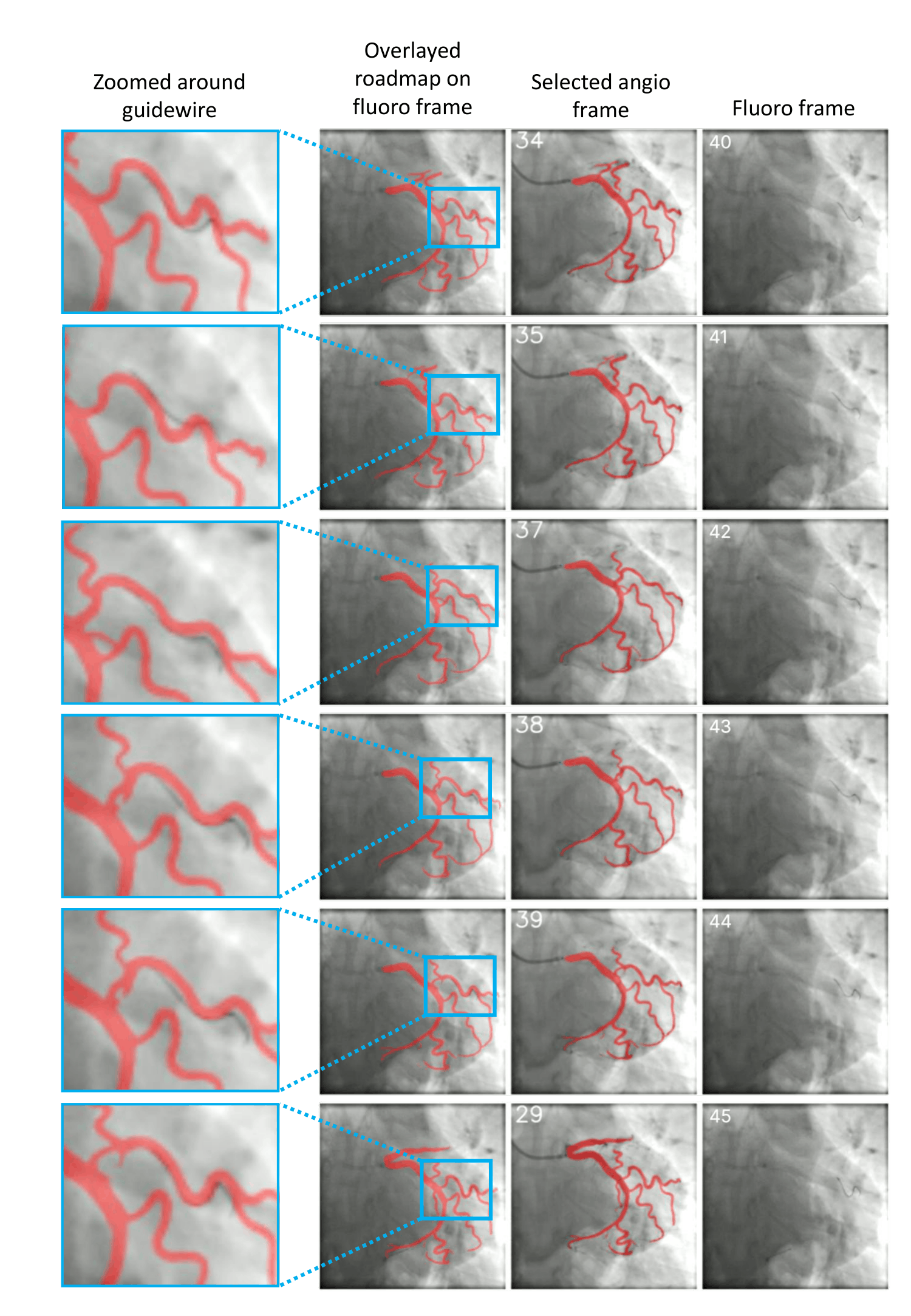}
    \caption{Qualitative result of our model on a few consecutive frames of an example angiography-fluoroscopy pair. The zoomed version shows the overlap of the guidewire with the predicted roadmap. The structural similarity of the shape of the guidewire and the roadmap signifies accurate cardiac phase matching. The numbers on top of frame depicts the frame number.}
    \label{fig:qual01}
\end{figure*}
The advantage that arises from a unified model is greater speed than having separate models for different tasks. Comparison of distance error and speed of different models under a two stage setup and our unified setup is shown in Fig.~\ref{fig:speed}. The two stage setup performs catheter tip tracking and cardiac phase matching sequentially using separate models, whereas the unified setup performs both tasks jointly within a single model. As shown, unifying the pipeline consistently improves computational efficiency across all backbones, with the most pronounced gains observed for STPM. While the two-stage STPM already achieves the lowest distance error among the evaluated methods, its unified counterpart further reduces inference latency without sacrificing accuracy, demonstrating the effectiveness of integrating spatial and temporal reasoning within a single model.

It is also worth noting that although CNN-based models are typically faster than transformer-based approaches, their advantage is substantially diminished in the conventional two-stage dynamic coronary roadmapping pipeline, where overall throughput is reduced to approximately 20 fps due to sequential processing. In contrast, the single-stage unified formulation removes this bottleneck and allows the CNN-based model to operate at up to 84 fps, better reflecting the inherent efficiency of convolutional architectures.

Unified STPM (ours) occupies a favorable region of the speed–accuracy trade-off, achieving the lowest error overall while operating at substantially higher throughput than its two-stage formulation and other unified baselines. This result highlights the complementary nature of our contributions: STPM provides a robust spatio-temporal matching mechanism, while the unified design removes redundant computation and enables efficient end-to-end inference. Together, these design choices yield a model that is both accurate and practical for real-time distance estimation, aligning with the objectives outlined throughout this work.

\subsection{Qualitative results}

\begin{figure}
  \centering
    \includegraphics[width=0.49\textwidth]{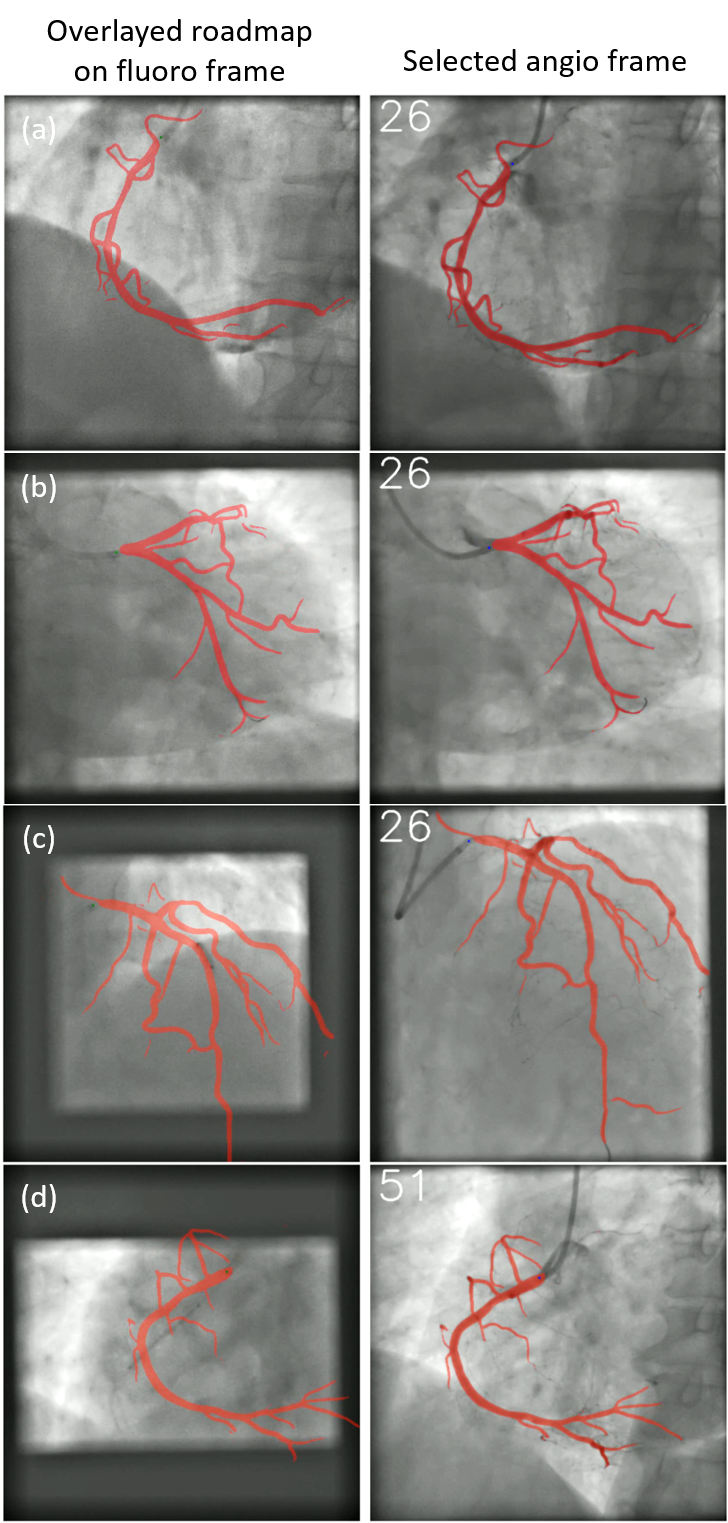}
    \caption{Example of DRM results on frames randomly selected from some challenging cases. (a) to (d) refers to 4 different angiography-fluoroscopy sequence pairs.}
    \label{fig:qual02}
\end{figure}

Figure \ref{fig:qual01} presents qualitative results of the proposed method on consecutive frames from an angiography fluoroscopy pair. The overlaid roadmap closely follows the guidewire trajectory across frames, indicating accurate cardiac phase matching. In particular, the structural alignment between the guidewire in fluoroscopy and the vessel tree from angiography remains consistent despite cardiac motion. The temporal consistency across frames suggests that the model effectively captures the underlying cardiac dynamics rather than relying on frame-wise appearance cues. The zoomed-in regions further highlight that fine vessel structures are well aligned with the guidewire, demonstrating that the model preserves spatial details while maintaining temporal coherence. This behavior is essential for reliable dynamic coronary roadmapping, where even small misalignments can affect clinical usability.

Figure~\ref{fig:qual02} shows qualitative results on more challenging cases. In Fig.~\ref{fig:qual02}(a) and Fig.~\ref{fig:qual02}(b), the guidewire or stent is fully occluded by the overlaid vessel branches. This is expected in accurate dynamic coronary roadmapping, where the projected vessel structure should coincide with the underlying device. The consistent overlap between the device and the vessel tree indicates correct cardiac phase matching and precise spatial alignment between angiography and fluoroscopy.

In Fig.~\ref{fig:qual02}(c) and Fig.~\ref{fig:qual02}(d), the fluoroscopy frames exhibit lower spatial resolution compared to the angiography reference, resulting in noticeable discrepancies in spatial appearance between the two modalities. In addition, fluoroscopy provides limited structural information, primarily from background anatomy and interventional devices such as the catheter or guidewire. This reduced and indirect set of cues makes phase matching more challenging. Despite this, the model achieves good alignment by leveraging spatio-temporal cues and capturing the underlying cardiac motion dynamics, leading to accurate phase matching even under such appearance variations. Fig.~\ref{fig:qual02}(d) also illustrates a case with a slight angulation change between angiography and fluoroscopy, which can be observed from the image borders. This geometric discrepancy introduces additional misalignment in the overlay. Overall, these examples highlight the robustness of the proposed method under occlusions, resolution differences, and mild geometric inconsistencies, while also illustrating remaining limitations in such scenarios.

\section{Conclusion}

In this work, we presented a unified and fully automated framework for ECG-free dynamic coronary roadmapping that jointly addresses cardiac phase matching and catheter tip tracking. By leveraging a large-scale spatio-temporal encoder pretrained on millions of unlabeled angiography and fluoroscopy frames, the proposed method effectively captures fine-grained cardiac motion dynamics that are essential for accurate temporal alignment. Unlike prior approaches that rely heavily on extensive frame-level dense annotations, our framework incorporates lightweight auxiliary tasks, including ECG R-peak detection and catheter tip tracking, to stabilize training and provide meaningful motion cues without requiring additional manual supervision. While existing methods typically achieve strong performance through extensive manual annotations, our approach instead relies on large-scale unlabeled data to learn a foundation model that not only reduces annotation requirements for dynamic coronary roadmapping, but also has the potential to generalize to a broader range of angiography-based tasks.

Comprehensive evaluations on clinical X-ray datasets demonstrate that the proposed approach achieves state-of-the-art performance in cardiac phase matching, yielding low temporal misalignment with high consistency across diverse imaging conditions. The results further highlight the importance of domain-specific spatio-temporal pretraining, as well as the benefit of jointly modeling spatial and temporal information within a unified architecture. In addition, the majority-voting post-processing strategy improves robustness during inference and naturally provides a confidence score that correlates well with phase-matching error, enabling uncertainty-aware deployment in clinical settings.

We further showed that the unified model achieves competitive catheter tip tracking performance under automatic initialization, while eliminating the need for separate task-specific models. Although a modest performance gap is observed compared to methods trained with manual initialization, this difference is not statistically significant and is outweighed by the advantages of reduced system complexity, faster inference, and improved suitability for fully automated DRM workflows.

Despite these promising results, several limitations remain and point toward directions for future work. In particular, while the proposed phase-matching approach achieves low average error, a non-negligible number of outliers persist in challenging cases. Achieving a fully reliable system may therefore require additional robustness, which could be obtained through increased annotated training set or by introducing further auxiliary tasks that provide complementary motion cues, such as guidewire detection. Moreover, although catheter tip tracking errors are low, closing the remaining gap to methods such as HiFT with manual initialization within a unified model remains an open challenge. Addressing this may require more advanced optimization strategies tailored for multi-task learning, enabling better task balancing and more effective feature sharing across objectives.

Overall, the proposed framework represents a step toward practical, real-time dynamic coronary roadmapping with minimal reliance on manual annotations or external signals. By demonstrating that large-scale spatio-temporal pretraining combined with carefully designed auxiliary tasks can effectively reduce the dependence on extensive labeled data, this work opens avenues for extending the learned representations to other interventional imaging tasks where motion understanding and robustness are critical.

\subsection*{Disclaimer}
The concepts and information presented in this paper are based on research results that are not commercially available. Future commercial availability cannot be guaranteed.










\bibliographystyle{cas-model2-names}

\bibliography{cas-refs}

@article{khan2022percutaneous,
  title={Percutaneous coronary intervention},
  author={Khan, Sohail Q and Ludman, Peter F},
  journal={Medicine},
  volume={50},
  number={7},
  pages={437--444},
  year={2022},
  publisher={Elsevier}
}

@article{piayda2018dynamic,
  title={Dynamic coronary roadmapping during percutaneous coronary intervention: a feasibility study},
  author={Piayda, Kerstin and Kleinebrecht, Laura and Afzal, Shazia and Bullens, Roland and Ter Horst, Iris and Polzin, Amin and Veulemans, Verena and Dannenberg, Lisa and Wimmer, Anna Christina and Jung, Christian and others},
  journal={European journal of medical research},
  volume={23},
  number={1},
  pages={36},
  year={2018},
  publisher={Springer}
}

@article{tehrani2013contrast,
  title={Contrast-induced acute kidney injury following PCI},
  author={Tehrani, Shana and Laing, Chris and Yellon, Derek M and Hausenloy, Derek J},
  journal={European journal of clinical investigation},
  volume={43},
  number={5},
  pages={483--490},
  year={2013},
  publisher={Wiley Online Library}
}

@misc{elion1989dynamic,
  title={Dynamic coronary roadmapping},
  author={Elion, Jonathan L},
  year={1989},
  month=oct # "~31",
  publisher={Google Patents},
  note={US Patent 4,878,115}
}

@inproceedings{zhu2010image,
  title={Image-based respiratory motion compensation for fluoroscopic coronary roadmapping},
  author={Zhu, Ying and Tsin, Yanghai and Sundar, Hari and Sauer, Frank},
  booktitle={International Conference on Medical Image Computing and Computer-Assisted Intervention},
  pages={287--294},
  year={2010},
  organization={Springer}
}

@inproceedings{manhart2011self,
  title={Self-assessing image-based respiratory motion compensation for fluoroscopic coronary roadmapping},
  author={Manhart, Michael and Zhu, Ying and Vitanovski, Dime},
  booktitle={2011 IEEE International Symposium on Biomedical Imaging: From Nano to Macro},
  pages={1065--1069},
  year={2011},
  organization={IEEE}
}

@article{kim2018registration,
  title={Registration of angiographic image on real-time fluoroscopic image for image-guided percutaneous coronary intervention},
  author={Kim, Dongkue and Park, Sangsoo and Jeong, Myung Ho and Ryu, Jeha},
  journal={International journal of computer assisted radiology and surgery},
  volume={13},
  number={2},
  pages={203--213},
  year={2018},
  publisher={Springer}
}

@article{ma2020dynamic,
  title={Dynamic coronary roadmapping via catheter tip tracking in X-ray fluoroscopy with deep learning based Bayesian filtering},
  author={Ma, Hua and Smal, Ihor and Daemen, Joost and van Walsum, Theo},
  journal={Medical image analysis},
  volume={61},
  pages={101634},
  year={2020},
  publisher={Elsevier}
}

@inproceedings{liu2024auxiliary,
  title={Auxiliary Input in Training: Incorporating Catheter Features into Deep Learning Models for ECG-Free Dynamic Coronary Roadmapping},
  author={Liu, Yikang and Zhao, Lin and Chen, Eric Z and Chen, Xiao and Chen, Terrence and Sun, Shanhui},
  booktitle={International Conference on Medical Image Computing and Computer-Assisted Intervention},
  pages={67--77},
  year={2024},
  organization={Springer}
}

@article{shechter2005prospective,
  title={Prospective motion correction of X-ray images for coronary interventions},
  author={Shechter, Guy and Shechter, Barak and Resar, Jon R and Beyar, Rafael},
  journal={IEEE transactions on medical imaging},
  volume={24},
  number={4},
  pages={441--450},
  year={2005},
  publisher={IEEE}
}

@article{timinger2005motion,
  title={Motion compensated coronary interventional navigation by means of diaphragm tracking and elastic motion models},
  author={Timinger, Holger and Krueger, Sascha and Dietmayer, Klaus and Borgert, Joern},
  journal={Physics in Medicine \& Biology},
  volume={50},
  number={3},
  pages={491},
  year={2005},
  publisher={IOP Publishing}
}

@article{faranesh2013integration,
  title={Integration of cardiac and respiratory motion into MRI roadmaps fused with x-ray},
  author={Faranesh, Anthony Z and Kellman, Peter and Ratnayaka, Kanishka and Lederman, Robert J},
  journal={Medical physics},
  volume={40},
  number={3},
  pages={032302},
  year={2013},
  publisher={Wiley Online Library}
}

@article{fischer2017mr,
  title={An MR-based model for cardio-respiratory motion compensation of overlays in X-ray fluoroscopy},
  author={Fischer, Peter and Faranesh, Anthony and Pohl, Thomas and Maier, Andreas and Rogers, Toby and Ratnayaka, Kanishka and Lederman, Robert and Hornegger, Joachim},
  journal={IEEE transactions on medical imaging},
  volume={37},
  number={1},
  pages={47--60},
  year={2017},
  publisher={IEEE}
}

@inproceedings{schneider2010model,
  title={Model-based respiratory motion compensation for image-guided cardiac interventions},
  author={Schneider, Matthias and Sundar, Hari and Liao, Rui and Hornegger, Joachim and Xu, Chenyang},
  booktitle={2010 IEEE Computer Society Conference on Computer Vision and Pattern Recognition},
  pages={2948--2954},
  year={2010},
  organization={IEEE}
}

@article{king2009subject,
  title={A subject-specific technique for respiratory motion correction in image-guided cardiac catheterisation procedures},
  author={King, Andrew P and Boubertakh, Redha and Rhode, Kawal S and Ma, YingLiang and Chinchapatnam, Phani and Gao, Gang and Tangcharoen, T and Ginks, Matthew and Cooklin, Michael and Gill, Jaswinder S and others},
  journal={Medical Image Analysis},
  volume={13},
  number={3},
  pages={419--431},
  year={2009},
  publisher={Elsevier}
}

@article{peressutti2013novel,
  title={A novel Bayesian respiratory motion model to estimate and resolve uncertainty in image-guided cardiac interventions},
  author={Peressutti, Devis and Penney, Graeme P and Housden, R James and Kolbitsch, Christoph and Gomez, Alberto and Rijkhorst, Erik-Jan and Barratt, Dean C and Rhode, Kawal S and King, Andrew P},
  journal={Medical image analysis},
  volume={17},
  number={4},
  pages={488--502},
  year={2013},
  publisher={Elsevier}
}

@inproceedings{demoustier2023contrack,
  title={Contrack: contextual transformer for device tracking in x-ray},
  author={Demoustier, Marc and Zhang, Yue and Narasimha Murthy, Venkatesh and Ghesu, Florin C and Comaniciu, Dorin},
  booktitle={International Conference on Medical Image Computing and Computer-Assisted Intervention},
  pages={679--688},
  year={2023},
  organization={Springer}
}

@article{islam2024self,
  title={Self-supervised learning for interventional image analytics: toward robust device trackers},
  author={Islam, Saahil and Murthy, Venkatesh N and Neumann, Dominik and Das, Badhan Kumar and Sharma, Puneet and Maier, Andreas and Comaniciu, Dorin and Ghesu, Florin C},
  journal={Journal of Medical Imaging},
  volume={11},
  number={3},
  pages={035001--035001},
  year={2024},
  publisher={Society of Photo-Optical Instrumentation Engineers}
}

@inproceedings{islam2024novel,
  title={A Novel Tracking Framework for Devices in X-ray Leveraging Supplementary Cue-Driven Self-supervised Features},
  author={Islam, Saahil and Murthy, Venkatesh N and Neumann, Dominik and Cimen, Serkan and Sharma, Puneet and Maier, Andreas and Comaniciu, Dorin and Ghesu, Florin C},
  booktitle={International Conference on Medical Image Computing and Computer-Assisted Intervention},
  pages={25--34},
  year={2024},
  organization={Springer}
}

@article{ravi2024sam,
  title={Sam 2: Segment anything in images and videos},
  author={Ravi, Nikhila and Gabeur, Valentin and Hu, Yuan-Ting and Hu, Ronghang and Ryali, Chaitanya and Ma, Tengyu and Khedr, Haitham and R{\"a}dle, Roman and Rolland, Chloe and Gustafson, Laura and others},
  journal={arXiv preprint arXiv:2408.00714},
  year={2024}
}

@article{tong2022videomae,
  title={Videomae: Masked autoencoders are data-efficient learners for self-supervised video pre-training},
  author={Tong, Zhan and Song, Yibing and Wang, Jue and Wang, Limin},
  journal={Advances in neural information processing systems},
  volume={35},
  pages={10078--10093},
  year={2022}
}

@inproceedings{wang2023videomae,
  title={Videomae v2: Scaling video masked autoencoders with dual masking},
  author={Wang, Limin and Huang, Bingkun and Zhao, Zhiyu and Tong, Zhan and He, Yinan and Wang, Yi and Wang, Yali and Qiao, Yu},
  booktitle={Proceedings of the IEEE/CVF conference on computer vision and pattern recognition},
  pages={14549--14560},
  year={2023}
}

@article{bardes2024revisiting,
  title={Revisiting feature prediction for learning visual representations from video},
  author={Bardes, Adrien and Garrido, Quentin and Ponce, Jean and Chen, Xinlei and Rabbat, Michael and LeCun, Yann and Assran, Mahmoud and Ballas, Nicolas},
  journal={arXiv preprint arXiv:2404.08471},
  year={2024}
}

@inproceedings{yan2021learning,
  title={Learning spatio-temporal transformer for visual tracking},
  author={Yan, Bin and Peng, Houwen and Fu, Jianlong and Wang, Dong and Lu, Huchuan},
  booktitle={Proceedings of the IEEE/CVF international conference on computer vision},
  pages={10448--10457},
  year={2021}
}

@inproceedings{cui2022mixformer,
  title={Mixformer: End-to-end tracking with iterative mixed attention},
  author={Cui, Yutao and Jiang, Cheng and Wang, Limin and Wu, Gangshan},
  booktitle={Proceedings of the IEEE/CVF conference on computer vision and pattern recognition},
  pages={13608--13618},
  year={2022}
}

@inproceedings{li2018high,
  title={High performance visual tracking with siamese region proposal network},
  author={Li, Bo and Yan, Junjie and Wu, Wei and Zhu, Zheng and Hu, Xiaolin},
  booktitle={Proceedings of the IEEE conference on computer vision and pattern recognition},
  pages={8971--8980},
  year={2018}
}

@article{assran2025v,
  title={V-jepa 2: Self-supervised video models enable understanding, prediction and planning},
  author={Assran, Mido and Bardes, Adrien and Fan, David and Garrido, Quentin and Howes, Russell and Muckley, Matthew and Rizvi, Ammar and Roberts, Claire and Sinha, Koustuv and Zholus, Artem and others},
  journal={arXiv preprint arXiv:2506.09985},
  year={2025}
}

@inproceedings{schroff2015facenet,
  title={Facenet: A unified embedding for face recognition and clustering},
  author={Schroff, Florian and Kalenichenko, Dmitry and Philbin, James},
  booktitle={Proceedings of the IEEE conference on computer vision and pattern recognition},
  pages={815--823},
  year={2015}
}

@article{khosla2020supervised,
  title={Supervised contrastive learning},
  author={Khosla, Prannay and Teterwak, Piotr and Wang, Chen and Sarna, Aaron and Tian, Yonglong and Isola, Phillip and Maschinot, Aaron and Liu, Ce and Krishnan, Dilip},
  journal={Advances in neural information processing systems},
  volume={33},
  pages={18661--18673},
  year={2020}
}

@inproceedings{chen2020simple,
  title={A simple framework for contrastive learning of visual representations},
  author={Chen, Ting and Kornblith, Simon and Norouzi, Mohammad and Hinton, Geoffrey},
  booktitle={International conference on machine learning},
  pages={1597--1607},
  year={2020},
  organization={PmLR}
}

@inproceedings{ghojogh2020fisher,
  title={Fisher discriminant triplet and contrastive losses for training siamese networks},
  author={Ghojogh, Benyamin and Sikaroudi, Milad and Shafiei, Sobhan and Tizhoosh, Hamid R and Karray, Fakhri and Crowley, Mark},
  booktitle={2020 international joint conference on neural networks (IJCNN)},
  pages={1--7},
  year={2020},
  organization={IEEE}
}

@inproceedings{he2020momentum,
  title={Momentum contrast for unsupervised visual representation learning},
  author={He, Kaiming and Fan, Haoqi and Wu, Yuxin and Xie, Saining and Girshick, Ross},
  booktitle={Proceedings of the IEEE/CVF conference on computer vision and pattern recognition},
  pages={9729--9738},
  year={2020}
}

@article{caron2020unsupervised,
  title={Unsupervised learning of visual features by contrasting cluster assignments},
  author={Caron, Mathilde and Misra, Ishan and Mairal, Julien and Goyal, Priya and Bojanowski, Piotr and Joulin, Armand},
  journal={Advances in neural information processing systems},
  volume={33},
  pages={9912--9924},
  year={2020}
}

@INPROCEEDINGS{Hadsell1640964,
  author={Hadsell, R. and Chopra, S. and LeCun, Y.},
  booktitle={2006 IEEE Computer Society Conference on Computer Vision and Pattern Recognition (CVPR'06)}, 
  title={Dimensionality Reduction by Learning an Invariant Mapping}, 
  year={2006},
  volume={2},
  number={},
  pages={1735-1742},
  doi={10.1109/CVPR.2006.100}}

@inproceedings{zhao2019weakly,
  title={A weakly supervised adaptive triplet loss for deep metric learning},
  author={Zhao, Xiaonan and Qi, Huan and Luo, Rui and Davis, Larry},
  booktitle={Proceedings of the IEEE/CVF International Conference on Computer Vision Workshops},
  pages={0--0},
  year={2019}
}

@article{oquab2023dinov2,
  title={Dinov2: Learning robust visual features without supervision},
  author={Oquab, Maxime and Darcet, Timoth{\'e}e and Moutakanni, Th{\'e}o and Vo, Huy and Szafraniec, Marc and Khalidov, Vasil and Fernandez, Pierre and Haziza, Daniel and Massa, Francisco and El-Nouby, Alaaeldin and others},
  journal={arXiv preprint arXiv:2304.07193},
  year={2023}
}

@inproceedings{caron2021emerging,
  title={Emerging properties in self-supervised vision transformers},
  author={Caron, Mathilde and Touvron, Hugo and Misra, Ishan and J{\'e}gou, Herv{\'e} and Mairal, Julien and Bojanowski, Piotr and Joulin, Armand},
  booktitle={Proceedings of the IEEE/CVF international conference on computer vision},
  pages={9650--9660},
  year={2021}
}



\end{document}